\documentstyle[preprint,aps]{revtex}
\begin{document}

 \newcommand \be {\begin{equation}}
\newcommand \ee {\end{equation}}
 \newcommand \ba {\begin{eqnarray}}
\newcommand \ea {\end{eqnarray}}
\def\oppropto{\mathop{\propto}}
\def\operarrow{\mathop{\longrightarrow}}
\def\opsimeq{\mathop{\simeq}}
\def\opeq{\mathop{=}}
\def \P{ {P(E,t)}}
\def \sinc{ \mbox{sinc}}
\def \Pe{ {P_{\rm eq}(E)}}
\def \om{ {\omega_{\rm eq} (\beta)}}

\title{\bf MODELS OF TRAPS AND GLASS PHENOMENOLOGY}

\vskip 3 true cm

\author{C\'ecile Monthus$^1$ \ and Jean-Philippe Bouchaud$^2$}

\address{\it $^1$  Service de Physique Th\'eorique, 
Centre d'\'etudes de Saclay,\\ Orme des Merisiers, 
91191 Gif-sur-Yvette C\'edex, FRANCE \\
$^2$ Service de Physique de l'\'Etat Condens\'e,
 Centre d'\'etudes de Saclay, \\ Orme des Merisiers, 
91191 Gif-sur-Yvette C\'edex, FRANCE \\ }

\vskip 1 true cm

%\date{\today}

\maketitle

\begin{abstract}
{We study various models of independent particles hopping between
 energy `traps' with a density of energy barriers $\rho(E)$, on a $d$ dimensional
lattice or on a fully connected lattice. If $\rho(E)$ decays exponentially, a true
dynamical phase transition between a high temperature `liquid' phase and a low
temperature `aging' phase occurs. More generally, however, one expects that for a large
class of $\rho(E)$, `interrupted' aging effects appear at low enough temperatures,
with an ergodic time growing faster than exponentially. The relaxation functions
exhibit a characteristic shoulder, which can be fitted
as stretched exponentials.  A simple way of  introducing interactions between the
particles leads to a modified model with an effective diffusion constant in energy
space, which we discuss in detail.}
\end{abstract}

\vfill

\noindent Electronic addresses : \hfill \break
monthus@amoco.saclay.cea.fr   \hfill \break
bouchaud@amoco.saclay.cea.fr

\vskip 1 true cm

\noindent  \hfill December 1995

\newpage

%\baselineskip 6mm
%\narrowtext

\section {Introduction}

Many very different glass formers exhibit surprisingly similar properties. 
For example, a common experimental feature is the `shouldering' of the relaxation
laws as the temperature is decreased \cite{Gotze,Science}. More precisely, the 
relaxation of the density
fluctuations evolves from a simple Debye exponential at high temperatures (liquid) to
a two-step process at lower temperature, where the correlation function first decays
rather quickly to a `plateau' ($\beta$ relaxation), and then departs from this
plateau value on a much longer time scale $\tau(T)$. This relaxation time  $\tau(T)$ 
grows
extremely fast as the temperature is decreased, in any case faster than $\exp{\Delta
\over T}$.  A very successful description of this divergence is the Vogel-Fulcher
law: $\Gamma_0 \tau(T) \sim e^{\Delta \over T - T_0}$, where $\Gamma_0^{-1}$ is a  microscopic
time scale \cite{VF}. However, other functional forms, such as $\Gamma_0 \tau(T) \sim  
e^{({\Delta \over T})^2}$, give reasonable fits of the data  \cite{Bass,Gotze}. When
$\tau(T)$ becomes of the order of the typical experimental time scales (say, a day), the
system is conventionally called a `glass', which is thus inherently out of
equilibrium.

A remarkable and popular theory of dynamical processes in supercooled liquids
is the so-called Mode-Coupling Theory (MCT), developped by G\"otze and others
\cite{Gotze}. Starting from a family of schematic equations which include a
non-linear, retarded feedback of the density fluctuations, one can show that there
exists an `ideal glass' transition temperature $T_c$, below which the correlation
function does not decay to zero (`broken ergodicity'). A notable prediction of the
theory is the existence of the two regimes $\beta$ and $\alpha$ mentioned above, and
a power-law divergence of the `slow' time scale $\tau(T)$ as $(T-T_c)^{-\gamma}$.
However, a detailed comparison with the experiments \cite{Comments} shows that the
transition temperature $T_c$, if it exists, is much higher than the Vogel-Fulcher
temperature, leaving a whole temperature interval $[T_0,T_c]$  where MCT
predicts a partial freezing (i.e. the $\alpha$ regime disappears) while the
experimental relaxation time is still finite (and behaves \`a la Vogel-Fulcher): 
the evidence for a (smeared) critical temperature is thus not very compelling. 
A way to circumvent this difficulty could be, as recently proposed in \cite{us}, to
work deep below $T_c$, where an extension of the MCT could be unambiguously tested.

However, the physical status of the MCT is not yet very clear: the
MCT equations are formally identical to those describing 
some mean field, infinite dimensional models of spin-glasses \cite{Thiru,CuKu,us},
where the presence of {\it quenched} disorder is assumed from the start. Further
work should clarify the interpretation of these underlying `spin' degrees of
freedom, the importance of finite dimensionality effects (in particular to allow
the existence of activated processes, which are presumably absent in infinite
dimension) and the justification of introducing quenched disorder by hand, rather than
letting it be `self-induced' by the dynamics \cite{SIQD}.

In view of these difficulties, it is interesting to investigate in detail
alternative phenomenological descriptions. Among quite a few other ones 
\cite{Gotze,Ritort}, a particularly simple picture, advocated by many authors over the
years, is the following: each particle is in a `cage', i.e. a potential well created by
its neighbours, from which it can escape through thermal activation 
\cite{Bass,Vilgis,Odagaki1,Odagaki2,BCM95}. As the temperature is decreased, the
probability distribution of local trapping times becomes very broad, and this
naturally leads to a two-step relaxation, stretched exponential decays, and Cole-Cole
susceptibility spectra. However, two important facets of these `trap' models have not
been addressed previously in the context of glasses:

$\bullet$ At low enough temperatures, the relaxation times exceeds the experimental
time scales, and aging effects become important \cite{Bou92,Struik,rev}. These aging effects have
been recently discussed within the framework of the MCT \cite{us,Franz}. Special
attention will be devoted to situations where aging is `interrupted' \cite{Bou92,BVH}
beyond a finite, but very long, time scale -- which is often the case experimentally.

$\bullet$ Since a priori all particles can move, the (random) potential well trapping
any one of them is in fact not quenched but time dependent, further enhancing the
probability of moving. In order to understand the glass transition, one must describe
how, in a self-consistent way, all motion ceases. This problem is similar to the one
alluded to above in the context of MCT. 

It is the aim of the present paper to discuss them in detail (a preliminary account of
this work can be found in \cite{BCM95}), as well as to investigate systematically other aspects of
these trap models, such as finite time properties. We also compare the results of
this approach to the observed properties of glasses, and emphasize its strengths and
weaknesses.

\section {Mean-field model of non-interacting traps}

\subsection {Definition of the model}
 
We first consider the thermal dynamics of {\it independent} particles in a space of
traps characterized by a given probability distribution $\rho(E)$ 
for the depth $E \in [0,+\infty[$ of traps \cite{Bou92}.
At temperature $T \equiv \beta^{-1}$, each particle may escape from its trap 
of depth $E$
 with rate $\Gamma_0 e^{-\beta E}$ per unit time. In the simplest
`mean-field' version, the particle chooses a new trap of depth $E'$ with probability
$\rho(E')$, but with no reference to any {\it spatial} structure: this will be
considered below. The probability density $P(E,t)$ to be in a trap of depth $E$ at
time $t$ therefore evolves in time according to the following equation  \be
{1 \over \Gamma_0} {\partial \P \over \partial t} = - e^{-\beta E} \ \P + \omega(t) \rho(E)  
\label{MEtrap}
\ee
where $\omega(t)=\int_0^{\infty} dE \ e^{-\beta E} \ \P$ 
is the average hopping rate at time t.
This equation has to be supplemented by some initial condition $P(E,t=0)=P_0(E)$.

A natural definition of the (two-time) correlation function within such a model is
the following:
\be
C(t_w+t,t_w) = \Pi(t_w+t,t_w) + q_0 [1- \Pi(t_w+t,t_w)] \label{Ctt}
\ee
where $\Pi(t_w+t,t_w)$ is the probability that the particle has not changed trap
between $t_w$ and $t_w+t$:
\be
\Pi(t_w+t,t_w) = \int_0^\infty dE \ P(E,t_w)
\ \exp[- \left(\Gamma_0 e^{-\beta E}\right) t]\label{Pi}
\ee
 $q_0$
is a certain number measuring the correlation between traps, which we shall set to
zero for simplicity. Eq. (\ref{Ctt}) assumes that no dynamics take place `inside a
trap' since $C \equiv 1$ until the particle has jumped out of the trap. If a stationary distribution can be reached at long times, the correlation function becomes {\it time translation invariant} and reads:
\be
\lim_{t_w \to \infty} C(t_w+t,t_w) = C_{\rm{eq}}(t) = 
\int_0^\infty dE \ \Pe \ \exp[- \left(\Gamma_0 e^{-\beta E}\right) t]\label{Ceq}
\ee
\subsection {Existence of a stationary distribution}

There exists a {\it normalisable} stationary distribution $\Pe$ at temperature
$T=\beta^{-1}$ only if \be
 \om \equiv \ {1 \over \displaystyle { \int_0^{+\infty} dE  \ e^{+\beta E} \ \rho(E)}} \ >0   
\label{omeq}
\ee
 in which case 
\be
\Pe = \om e^{+\beta E} \ \rho(E)    
\label{Peq}
\ee
In particular, $\Pe$ always exists at infinite temperature $(\beta=0)$ and is equal
to the `bare' distribution $\rho(E)$. The transition between the high temperature
phase ($T >T_0$) where $\Pe$ exists, and the low temperature phase $T <T_0$ where
$\Pe$ ceases to be normalisable, takes place at the temperature $T_0$ defined by
\be
{1 \over T_0} = \beta_0 \equiv \lim_{E \to \infty}  \left[-{ {\log\rho(E)} \over E} \right]
\label{defbeta0}
\ee 
We may therefore distinguish the three cases  \hfill \break
$\bullet$ $T_0=0 $ if $\rho(E)$ decays faster than exponentially at large $E$.
 \hfill \break
$\bullet$ $T_0=\infty $ if $\rho(E)$ decays slower than exponentially at large $E$.
 \hfill \break 
$\bullet$ $T_0$ is finite in the interesting case where
$\rho(E)$ decays exponentially as $e^{-\beta_0 E} $ at large $E$.  This exponential form is suggested by
the mean-field replica theory of spin-glasses \cite{MPV}, the random energy model
\cite{Derrida}, and by phenomenological arguments in the context of glasses
\cite{Odagaki2}. It has also been found exactly in the one-dimensional `random force' model \cite{RF} and was used to interpret anomalies in the transport properties of amorphous conductors (see e.g. \cite{BG90}). 
Its ubiquity in random systems is probably related to the fact that it is 
a stable extreme value distribution \cite{BM}.

For this exponential density of states, the trapping time $\tau \equiv \Gamma_0^{-1} e^{\beta E}$ is
distributed as a power law for large $\tau$:
\be
\Psi(\tau) \oppropto_{\tau \to \infty}  {\Gamma_0 \over (\Gamma_0 \tau)^{1+x}} \qquad x \equiv {T \over T_0}
\label{psitau}
\ee
Since this model exhibits a true finite temperature `glass' transition, its study 
is most interesting. However, the case of a Gaussian density of states, where strictly
speaking $T_0=0$, exhibits several features very similar to those observed in real
glasses, and will thus also be studied in detail.

\subsection{ Relaxation towards 
the equilibrium distribution for $T>T_0$.}

At infinite temperature $(\beta=0)$, the solution of (\ref{MEtrap}) may be directly
written since $\omega(t) \equiv1$
\be
\P = \rho(E) + \left[ P(E,0) - \rho(E) \right] e^{- \Gamma_0 t}  
\label{Relaxinf}
\ee

At finite temperature, we may analyse the solution of (\ref{MEtrap})
through a Laplace transform: introducing $\hat P(E,\lambda) \equiv \int_0^\infty dt
[P(E,t)-\Pe] e^{-\lambda \Gamma_0 t}$, one obtains:
 \be
\hat P(E,\lambda) = { { {1 \over \Gamma_0} \big[ P(E,t=0)-\Pe \big]+ 
\hat \omega(\lambda) \rho(E) }
\over {\lambda + e^{-\beta E}}}
\label{Relaxl} \ee
with the definition 
\be
\hat \omega(\lambda)\equiv \int_0^{\infty} dE' \  e^{-\beta E'}\hat P(E',\lambda) 
\label{normf}
\ee
Solving for $\hat \omega(\lambda)$, one finds:
\be
\hat \omega(\lambda)=  {1 \over {\Gamma_0 \lambda}} \bigg[ \omega_{\rm eq} - 
\frac{\int_0^\infty dE \ \frac{P(E,0)}{1 +
 \lambda e^{\beta E}}} {\int_0^\infty dE \ \frac {\rho(E) e^{\beta E} }
{1 + \lambda e^{\beta E}}} \bigg]
\label{hatomega}
\ee
In the interesting case of an exponential density of states $\rho(E) \propto
e^{-\beta_0 E}$, the denominator has a regular expansion up to
the order $\lambda^n$, where $n$ is the integer part of $x-1$ ($x = \frac{T}{T_0}$), 
followed by a singular term of
order $\lambda^{x-1}$. Hence, if $P(E,0)$ is sharply
peaked (e.g $P(E,0)=\delta(E-E_0)$), the small $\lambda$ expansion of $\hat P(E,\lambda)$ has a 
$\lambda^{x-2}$ singularity, indicating that 
the difference $P(E,t)-\Pe$ decays for large times as 
\be
P(E,t)-\Pe \oppropto_{t \to \infty} t^{-(x-1)}
\ee
 This decay is, not surprisingly, precisely the same as that of the correlation 
function
$C_{\rm {eq}}$ defined by Eq. (\ref{Ceq}) \cite{BCM95}
\be
C_{\rm {eq}} (t) =\int_0^\infty dE \  \om e^{+\beta E} \ \beta_0 e^{-\beta_0 E}  
 \ \exp[- \left(\Gamma_0 e^{-\beta E}\right) t] \opsimeq_{t \to \infty} \om
\Gamma(x) \ (\Gamma_0 t)^{-(x-1)}
\ee Note that the relaxation
time $\tau(T)$, defined as the time after which the correlation has decayed down 
to a certain
value $c$, diverges {\it \`a la} Vogel-Fulcher, i.e. 
\be
 \tau(T) \simeq {1 \over \Gamma_0} \bigg( {c \over {\om \Gamma(x) }} \bigg)^{\left({T_0 \over T-T_0}\right)}
\ee
Another usual definition of the `terminal' time, through $\tau_1 = \int_0^\infty dt \ C_{\rm {eq}}$,
would however lead to a power-law divergence {\it at a higher temperature} $2T_0$
\be
\tau_1 = {\om \beta_0 \over \Gamma_0} \int_0^{\infty} dE \ e^{(2 \beta - \beta_0 ) E} \opeq_{T > 2T_0}
{\om \over \Gamma_0} \ \left({T \over {T- 2T_0}}\right)
\ee 
The existence of {\it two} characteristic temperatures in this model was recently emphasized in the context of glasses by Odagaki \cite{Odagaki2}.
If $\rho(E)$ decays faster than exponentially (say as $\exp(-E^{1+\nu})$), the relaxation
towards equilibrium is faster than any power-law, since the small $\lambda$
expansion of $\hat P(E,\lambda)$ is regular. However, slower than exponential
relaxation (such as $\exp -(\log t)^{1+\nu}$) is expected. More generally, the
 long time behaviour
of
$P(E,t)-\Pe$ is the same as that of $C_{\rm {eq}}$ for large $t$.

\subsection {Non normalisability and aging in the low temperature phase.}
\label{sec-aging}

When there is no equilibrium distribution (\ref{omeq}), one expects that the
dynamics will never become stationary \cite{Bou92,BD95,Bardou,BCM95}, and this
 leads to
aging effects. One may look 
 at large time for a scaling
solution for the equation (\ref{MEtrap}). We introduce the natural
dimensionless scaling variable $u = {e^{\beta E} \over {\Gamma_0 t}}$ 
and a dimensionless function $\phi$ normalized as $\int_0^{\infty} du \  \phi(u) =1$
and try to find the asymptotic distribution function through the form 
\be
P(E,t) \opsimeq_{t \to \infty} \beta \ u \ \phi(u)
\label{scaling}
\ee
 The resulting equation for $\phi$ is
\be
 u^2 {d \phi \over du} +(u-1) \phi(u) = -{1 \over \beta}  \ \ 
\rho \left({ {\log\left( \Gamma_0 t u \right)} \over \beta} \right)  \ 
\int_{1 \over {\Gamma_0 t}}^{\infty} dv \ {\phi(v) \over v}
\label{eqphi}
\ee
The left-handside presents a $t$ independent limit as $t \to \infty$
only if the integral on the left-handside diverges
in such a way that it compensates exactly the decay of
 $\rho \left({ {\log\left( \Gamma_0 t u \right)} \over \beta} \right)$ 
as $t \to \infty$.

Consider for example the case of an exponential density of states
$\rho(E) = \beta_0 \ e^{- \beta_0 E}$, for which there is no equilibrium distribution
when $x \equiv {T \over T_0} \in ] 0,1[$.

In order to get some sensible limit of equation (\ref{eqphi}) as $t
\to \infty$, the function $\phi$ must present the singularity \be \phi(u) \opsimeq_{u
\to 0} \gamma \ u^{-x} \label{sing0}
\ee
where $a$ is some normalisation constant. 
The solution of the resulting equation
\be
 u^2 {d \phi \over du} +(u-1) \phi(u) = -\gamma \  u^{-x}
\label{eqphiexp}
\ee
 satisfying the normalization condition 
$\int_0^{\infty} du \ \phi(u)=1$ reads
\be
\phi(u) = { {\sin \pi x} \over {\pi \Gamma(x)}} \ {1 \over u} \
e^{ \displaystyle -{1 \over u} }
\int_0^{ \displaystyle {1 \over u}} dy \ y^{x-1}\  e^y 
\label{solphi}
\ee

The consequence of the scaling form (\ref{scaling}) on the correlation
function $C$ or $\Pi$ defined as the probability for the particle to remain
 in the same trap during the interval $[t_w,t_w+t]$ (see Eq. (\ref{Ctt}))
\be
\Pi(t_w+t,t_w ) = \int_0^{\infty} dE \ P(E,t_w) \ 
 e^{- \left( \Gamma_0 e^{- \beta E}\right) t}
\label{Pidef}
\ee
is the aging behavior \be \Pi(t_w,t_w+t ) \simeq  \int_0^{\infty} du
\ \phi(u) \ 
 e^{ \displaystyle - {1 \over u} \left( {t \over t_w} \right) } 
\label{Piaging}
\ee
where the two times only appear now through the combination 
$\left({t \over t_w}\right)$. This must be contrasted with the situation
prevailing for $T>T_0$, where as soon as $t_w \gg \tau(T)$, $P(E,t_w)$
ceases to depend on $t_w$ and the correlation function only depends on the time
difference $t$, and is plotted in Fig. 1-a for different values of $T \over T_0$
(We have included a `fast' relaxation inside each trap -- seen as a simple harmonic
 well, in order to mimic the $\beta$ regime of glasses). 

For the case of the exponential density of states, using equation (\ref{solphi})
we recover the explicit formula derived in \cite{BD95} through another approach
\be
\Pi(t_w+t,t_w ) \simeq  { {\sin \pi x} \over {\pi }}
\int_{ { \left({t \over t_w}\right) } \over {1+\left({t \over t_w}\right)Ê}}^{1} dv 
\left(1-v \right)^{x -1} v^{- x} 
\label{Piexp}
\ee
with  the asymptotic behaviors
\be
\Pi(t_w+t,t_w ) \simeq   
 1-   {\sinc [\pi (1-x)]}  
\left({t \over t_w}\right)^{1-x } \qquad \hbox{for} \ t \ll t_w  
 \label{agingexp0}
\ee    
\be
\Pi(t_w+t,t_w ) \simeq    {\sinc [\pi x]} 
\left({t \over t_w}\right)^{-x }   \qquad \hbox{for} \ t \gg t_w   \ 
 \label{agingexpinf}
\ee
where we have introduced the notation $\sinc [u] \equiv {\sin u \over u}$.

\subsection{Case of Gaussian density of states at low temperature : \hfill \break
stretched exponentials and interrupted aging}

Let us consider the interesting case of a Gaussian density of states
\be
\rho(E)={2 \over {\sqrt{\pi} E_c}} 
\ e^{ \displaystyle - \left( {E \over E_c} \right)^2}
\ee
which corresponds to a situation where the energy barriers results from a (weak)
interaction with many neighbours.
Strictly speaking, there is a stationary distribution for any positive
temperature (\ref{Peq})
 \be
\Pe = \om {2 \over {\sqrt{\pi} E_c}} \ e^{\displaystyle +\beta E} \ 
 e^{ \displaystyle - \left( {E \over E_c} \right)^2} \equiv {\cal N} \
e^{\displaystyle - { {(E-E^*)^2} \over E_c^2 }}  \qquad \mbox{where} \qquad E^*= {E_c\over 2} (\beta E_c)     \label{PeqG}
\ee
However, at very low temperature $T \ll E_c$, there exists an approximate
aging behavior on the time interval satisfying
\be
0 \ll \ln( \Gamma_0 t) \ll (\beta E_c)^2
\ee
but this aging phenomenon progressively disappears as time becomes large.
The physical reason is the following: as can be seen directly from $\rho(E)$, the distribution of relaxation times $\tau = \Gamma_0^{-1} e^{\beta E}$ can be written (for large $t$) as
\be
\Psi(\tau) \oppropto_{\tau \to \infty} \tau^{-1-{ {\ln(\Gamma_0 \tau)} \over {(\beta E_c)^2}}}
\label{PsiG}
\ee
Hence, the parameter $x ={T \over T_0}$ defined in the previous paragraphs 
is replaced by a slowly varying function  $x(t)=2{ {\log(\Gamma_0 t)} \over {(\beta
E_c)^2}}$. (The factor $2$ comes from a more careful analysis -- see below). The
above inequality thus corresponds to the case where $x(t) \ll 1$, where aging effects
are indeed expected.

More explicitly, one may look for an approximate scaling solution of (\ref{MEtrap})
analogous to (\ref{scaling}) 
\be
P(E,t) \simeq \beta \ u \ \phi_t(u)
\ee
where $u={e^{\beta E} \over {\Gamma_0 t}}$ and where the function $\phi_t$
 is supposed to vary slowly in time. More precisely, in the resulting equation for 
$\phi_t$
\be
-u t {{\partial \phi_t} \over {\partial t}}+
 u^2 { {\partial  \phi_t} \over {\partial u}} +(u-1) \phi_t(u) = -{1 \over \beta}  \ \ 
\rho \left({ {\log\left( \Gamma_0 t u \right)} \over \beta} \right)  \ 
\int_{1 }^{\infty} {dv \over v} \ \phi_t \left({v \over {\Gamma_0 t}}\right) 
\label{eqphit}
\ee
we assume that the first term
containing the time derivative is negligible in comparison to others.
We now therefore consider the simplified equation
\be
 u^2 {d \phi_t \over du} +(u-1) \phi_t(u) \simeq  -{1 \over \beta}  \ \ 
\rho \left({ {\log\left( \Gamma_0 t u \right)} \over \beta} \right)  \ 
\int_{1 }^{\infty} {dv \over v} \ \phi_t \left({v \over {\Gamma_0 t}}\right) 
\label{eqphits}
\ee
and we will discuss the validity of the hypothesis
 made on the explicit solution obtained for (\ref{eqphits}). 

For $\Gamma_0 t \gg 1$, the function $\phi_t$ must present
 the following asymptotic behavior
\be
\phi_t(u) \opsimeq_{u \to 0} A(t) \  u^{-x(t)} \ 
e^{ \displaystyle - \left({\log u \over { \beta E_c}}\right)^2}
\label{phit0}
\ee 
and (\ref{eqphits}) becomes
\be
 u^2 {d \phi_t \over du} +(u-1) \phi_t(u) \simeq
A(t) \  u^{-x(t)} \ 
e^{ \displaystyle - \left({\log u \over { \beta E_c}}\right)^2}
\ee 
The solution reads 
\be
\phi_t(u) = A(t) {1 \over u} \
e^{ \displaystyle -{1 \over u} }
\int_0^{ \displaystyle {1 \over u}} dy \ y^{x(t)-1}\ 
e^{ \displaystyle - \left({\log y \over { \beta E_c}}\right)^2} \  e^y 
\label{solphit}
\ee
where $A(t)$ is determined through the
normalisation condition $\int_0^{\infty} du \phi_t =1$
\be
{1 \over A(t)}=\int_0^{\infty} dy \ e^{-y} \ \int_{-\infty}^{+\infty} dz \
 { {e^{z x(t)}Ê} \over {e^z +y}} \ e^{-\left({z \over {\beta E_c}}\right)^2}
\label{At}
\ee
The asymptotic behaviors of $\phi_t(u)$ are respectively given by (\ref{phit0})
for $u \to 0$, and by the following for $u \to \infty$
\be
\phi_t(u) \opsimeq_{u \to \infty} {A(t)  \over x(t)}\  u^{-1-x(t)} \ 
e^{ \displaystyle - \left({\log u  \over { \beta E_c}}\right)^2}
\label{phitinfty}
\ee

We may now estimate the order of the term dropped in (\ref{eqphit})
\be
\left\vert u \  t \  {{\partial \phi_t} \over {\partial t}}  \right\vert =
\left\vert u \  t\  {d x(t) \over dt}\  
 \ { {\partial \phi_t} \over{\partial x(t)} } \right\vert
 = \left\vert {2 \over { (\beta E_c)^2}} \  u \ 
{ {\partial \phi_t} \over{\partial x(t)} } \right\vert \sim 
\left\vert {{2\ \log u} \over { (\beta E_c)^2}}  u \ \phi_t (u) \right\vert\ 
\ee
The comparison with other terms of (\ref{eqphit})
shows that the approximation (\ref{eqphits}) holds
 at low temperature $(\beta  E_c)^2 \gg 1$, and not too large
 $u$, i.e.: ${{\log u} \over { (\beta E_c)^2}} \ll  1$,

Let us now explain how our analysis is compatible
 with the relaxation towards the equilibrium distribution (\ref{PeqG}).
At low temperature ($\beta E_c \gg 1$), this
 equilibrium distribution is concentrated in a region of
scale $E_c$ around the central value $E^*=\frac{\beta E_c^2}{2}$ which is very large
compared to $E_c$. This is why during a very long time interval,
 the probability distribution $P(E,t)$ is pushed towards 
larger and larger energy as if there were no equilibrium distribution. 
More precisely, as soon as $\Gamma_0 t \gg 1$,   
 $P(E,t)$ enters an aging-like
scaling regime in the variable $u={e^{\beta E} \over {\Gamma_0 t}}$
as in the case of the exponential distribution below the transition ($T<T_0$).
From the above remark concerning the shape of $\Psi(\tau)$ (Eq. \ref{PsiG}), one expects 
that the
aging behaviour disappears  as $x(t)$ gets close to the value $1$. Then
the normalisability of $\Pe$ becomes noticeable, the probability distribution $P(E,t)$
begins to stabilize, and the scaling variable  $u={e^{\beta E} \over {\Gamma_0 t}}$
looses its meaning.
 In the limit $x(t) \to \infty$, the function $\phi_t$ becomes concentrated 
in the region $u \to 0$,
and the asymptotic behavior (\ref{phit0}) precisely corresponds 
to the equilibrium distribution (\ref{PeqG}) when one replaces $u$ by ${e^{\beta E}
\over {\Gamma_0 t}}$. 

We now turn to the correlation function $\Pi$ defined in (\ref{Pidef}).
For $\Gamma_0 t_w \gg 1$, one has in terms of the function $\phi_{t_w}(u)$ of equation
(\ref{solphit})
  \be
\Pi(t_w+t,t_w ) \simeq \int_0^{\infty} du \ \phi_{t_w}(u) \ 
 e^{\displaystyle  - {1 \over u} \left({ t \over t_w}\right)}
\ee
Here, contrary to the real aging behavior (\ref{Piaging}),
 $\Pi(t_w+t,t_w )$ is not a function of $\left({ t \over t_w}\right)$ only,
since the function $\phi_{t_w}$ slowly varies with the waiting time $t_w$.
In particular, using the asymptotic behaviors (\ref{phit0}) and (\ref{phitinfty}),
one has respectively at short time $t \ll t_w$  
\be
\Pi(t_w+t,t_w ) \opsimeq_{t \ll t_w}
1-  A(t_w) \int_0^{\infty} du \  \  u^{-x(t_w)} \ 
e^{ \displaystyle - \left({\log u \over { \beta E_c}}\right)^2} \ 
 \left( 1- e^{\displaystyle  - {1 \over u} \left({ t \over t_w}\right)} \right)
\label{PiG0}
\ee
and at large time $t \gg t_w$
\be
\Pi(t_w+t,t_w )  \opsimeq_{t \gg t_w}
 {A(t_w)  \over x(t_w)} \int_0^{\infty} dy 
\ y^{x(t_w)-1} \ e^{ \displaystyle - \left({\log y \over { \beta E_c}}\right)^2}
 e^{\displaystyle  - y\left({ t \over t_w}\right)}
\label{PiGinfty}
\ee 

To get more explicit behaviors, we specialize to the interesting regime
 $0 <x(t_w)  <1$ where the aging-like behavior occurs.
In this regime, the normalization constant (\ref{At}) may be expanded
according to
\be
{1 \over A(t_w)}=  \bigg(1- {1 \over {(\beta E_c)^2}} \  
 {\partial^2 \over {\partial x(t_w)^2} }+ \cdots \bigg)
\int_0^{\infty} dy \ e^{-y} \ \int_{0}^{+\infty} dz \
 { {z^{ x(t_w)}} \over {z +y}} 
\ee
\be  =
 e^{\displaystyle -  \left({1 \over {\beta E_c}}
 {\partial \over {\partial x(t_w)}}\right)^2 } \ \bigg[
{{\pi \Gamma(x(t_w))} \over {\sin \pi x(t_w)}}\bigg]
\ee

The explicit expression of the first correction in the small parameter
 ${1 \over {(\beta E_c)^2}}$
%\be  
%{1 \over A(t_w)}= {{\pi \Gamma(x(t_w))} \over {\sin \pi x(t_w)}}
 %\bigg[ 1 - {1 \over {(\beta E_c)^2}} 
%\bigg({ {\Gamma''(x(t_w))} \over {\Gamma(x(t_w))}} -2 \pi
% { {\Gamma'(x(t_w))} \over {\Gamma(x(t_w))}}
%{{ \cos\pi x(t_w)} \over {\sin \pi x(t_w)}} 
%+ \pi^2  \left( {{1+ \cos^2\pi x(t_w)} \over {\sin^2 \pi x(t_w)}}  \right) %\bigg) + 
%\cdots \bigg]
%\ee
shows that the approximation 
\be 
A(t_w) \simeq {{\sin \pi x(t_w) } \over {\pi \Gamma(x(t_w))}} 
\ee
is valid for 
${1 \over {\beta E_c}} \ll x(t_w) \ll 1- {1 \over {\beta E_c}}$.

The two asymptotic expressions (\ref{PiG0}-\ref{PiGinfty}) may be analysed 
in the same way. At short times
\be
\Pi(t_w+t,t_w ) \opsimeq_{t \ll t_w}
1-  A(t_w) \ e^{\displaystyle -  \left( {1 \over {\beta E_c}}
 {\partial \over {\partial x(t_w)}}\right)^2 } \ 
\bigg[ { {\Gamma(x(t_w))} \over {1-x(t_w) }} \ 
 \left({ t \over t_w} \right)^{1-x(t_w) } \bigg]
\ee
can be approximated by
\be
\Pi(t_w+t,t_w ) \opsimeq_{t \ll t_w}
1- 
 {\sinc [\pi  (1-x(t_w))}]  \ 
 \left({ t \over t_w} \right)^{1-{x( t_w)} } \label{Pi1}
%\bigg[ 1+ {1 \over {(\beta E_c)^2}} \bigg( 
%2 { {\Gamma'(x(t_w))} \over {\Gamma(x(t_w))}}
%\left( {1 \over {1-x(t_w)}}- \pi {{ \cos\pi x(t_w)} \over {\sin \pi %x(t_w)}}  \right) 
\ee
%\be
%+\pi^2  \left( {{1+ \cos^2\pi x(t_w)} \over {\sin^2 \pi x(t_w)}}\right) -{ {2} \over {(1-x(t_w))^2}}
% + \left(  { {\Gamma'(x(t_w))} \over {\Gamma(x(t_w))}} +
% {1 \over {1-x(t_w)}} \right)
%\log{ \tau \over t_w}  -\log^2{ \tau \over t_w} \bigg) + \cdots      \bigg]
%\ee
 for 
${1 \over {\beta E_c}} \ll x(t_w) \ll 1- {1 \over {\beta E_c}}$ and
for ${t \over t_w}$ not exponentially small in $(\beta E_c)^2$.

At large times,
\be
\Pi(t_w+t,t_w )  \opsimeq_{t \gg t_w}
 {A(t_w)  \over x(t_w)} \ e^{\displaystyle - \left( {1 \over {\beta E_c}}
 {\partial \over {\partial x(t_w)}}\right)^2 } \ 
\bigg[ \Gamma(x(t_w)) \left({ t \over t_w}\right)^{-x(t_w) }\bigg]
\ee 
may be reduced to
\be
\Pi(t_w+t,t_w ) \opsimeq_{t \gg t_w}
 {\sinc [\pi  x(t_w)]}  \ 
 \left({ t \over t_w} \right)^{-{x( t_w) } } \label{Pi2}
%\bigg[ 1+ {1 \over {(\beta E_c)^2}} 
%\bigg(2 { {\Gamma'(x(t_w))} \over {\Gamma(x(t_w))}}
%\left(\log{ \tau \over t_w}-  \pi {{ \cos\pi x(t_w)} \over {\sin \pi %x(t_w)}} \right)
\ee
%\be 
%+ \pi^2  \left( {{1+ \cos^2\pi x(t_w)} \over {\sin^2 \pix(t_w)}}\right)  
%-\log^2{ \tau \over t_w} \bigg) + \cdots                \bigg]
%\ee
 for 
${1 \over {\beta E_c}} \ll x(t_w) \ll 1- {1 \over {\beta E_c}}$ and
${ t \over t_w}$ not exponentially large in $(\beta E_c)^2$.

As anticipated, these expressions are very similar to those obtained for 
the exponential density of states, provided one defines an effective time dependent
parameter $x(t)$. [Actually, when $|\ln  \frac{t}{t_w}| \gg 1$, a slighly more
accurate expression of (\ref{Pi1},\ref{Pi2}) is obtained by replacing the exponent $x(t_w)$ by
$\frac{x(t)}{2}$, in agreement with the naive interpretation of Eq.(\ref{PsiG})]

Let us now concentrate on times such that
$x(t_w) > 1$: aging effects cease, and the stationary dynamics resume. The resulting
correlation function \be
C_{\rm {eq}}(t) =
\int_0^\infty dE 
\ \om \ e^{\beta E} {2 \over { \sqrt{\pi} E_c}} e^{- \left({E \over E_c}\right)^2} \
\exp[- \left(\Gamma_0 e^{-\beta E}\right) t] \ee
 is plotted in Fig. 1-b and displays very interesting features. 
To a very good approximation, one may replace the effect of the last exponential by a
simple cut-off:  
\be
C_{\rm {eq}}(t) \simeq
\int_{{1 \over \beta} \ln(\Gamma_0 t)}^\infty 
dE \ \om \ e^{\beta E} {2 \over { \sqrt{\pi} E_c}} e^{- \left({E \over E_c}\right)^2}
\ee
One thus finds the following behaviour:
\be
{\partial C_{\rm {eq}}(t) \over \partial \ln (\Gamma_0 t)} \simeq 
- {1 \over {\sqrt{\pi}  \beta E_c}} \ \ 
\ e^{ \displaystyle - \left( {\ln ({t \over \tau(T)}) \over \beta E_c} \right)^2}
\label{CG}
\ee
where $\tau(T) \equiv \frac{1}{\Gamma_0} \exp {(\beta E_c)^2 \over 2}$ is the time
scale  associated with the
$\alpha$ regime. As mentioned in the introduction, the experimental data on $\tau(T)$ 
can indeed be fitted by such a law \cite{Bass}. Note however that the `terminal' time
scale, defined  as $\tau_1 = \int_0^\infty C_{\rm {eq}} dt$, and extracted, in
supercooled liquids, from viscosity measurements, is given by $\tau_1 \simeq {1 \over
\Gamma_0} e^{{3 \over 4} (\beta E_c)^2 }$ at low temperature {\footnote{ Note however that the long time behaviour of $C_{\rm{eq}}(t)$ is
well approximated by $C_{\rm{eq}}(t) \propto ({\tau_2 \over t})^\mu$ with $\mu={({T \over E_c})^2 \log \Gamma_0 t}$ and $\tau_2 = \Gamma_0^{-1} 
\exp (\beta E_c)^2$, showing that there is actually quite a bit of
freedom to choose the rescaling time factor $\tau(T)$ without affecting
too much the quality of the scaling. }}.

Eq. (\ref{CG}) shows that the correlation function measured for different temperatures 
approximately fall onto a master curve if one plots them as a function of 
$t \over \tau(T)$, provided the variation of $\beta E_c$ (compared to that of
$\tau(T)$) can be neglected. Furthermore, as emphasized in \cite{Castaing}, the shape
of the relaxation function given by  Eq. (\ref{CG}) is actually very close, in the
region where $t$ and $\tau(T)$ are not too different, to a {\it stretched exponential} {\footnote{Note: the stretched exponential exponent is usually called $\beta$, although it describes the $\alpha$ relaxation in glasses, and 
should furthermore not be confused with the inverse temperature. We have thus proposed in \cite{us} to call it $\alpha$.}}
$C_{\rm {eq}}(t) = \exp -\left ({t \over \tau(T)} \right)^{\alpha(T)}$, provided one
identifies $\alpha(T) \simeq (1 + {(\beta E_c)^2 \over 2})^{-{1 \over 2}}$
\cite{Castaing}.

These features are again reminiscent of the experimental ones, where an approximate
 scaling of
$C_{\rm {eq}}(t)$ for various temperatures can indeed be achieved, and where a stretched 
exponential
(or Kohlrausch) form for the decay in the $\alpha$ regime is quite often proposed
 (see the 
in \cite{Gotze}). Interestingly, the exponent $\alpha$ extracted from experimental 
fits has a tendency to decrease as the temperature is decreased (see, e.g., Fig. 16 of
\cite{DPS}). 

Finally, let us stress that many of the results of the present section are not 
restricted to the
Gaussian form of the density of states, and would still hold if 
$\rho(E) \propto \exp -({E \over E_c})^{1+\nu}$, $\nu > 0$ with $(\beta E_c)^2$ is
replaced by  $(\beta E_c)^{1 + \frac{1}{\nu}}$ (up to $\nu$ dependent prefactors).
Such a generalized model has been recently discussed in the context of glasses in
\cite{Arkipov}. An interesting point to notice is that the variance $\sigma^2$ of the log-normal form of ${\partial C_{\rm{eq}}(t) \over \partial \ln t}$ is related to its maximum through $\sigma^2 = \frac{1}{\nu} \ln \tau(T)$. Such a relation was also obtained by Souletie using different arguments \cite{Souletie}, and is not incompatible with experimental data.

The conclusion of this section is that stretched exponential decay and scaling of 
the relaxation curves can occur in the absence of any criticality or cooperative
effects, and arises even when the distribution of local energy barriers decays quite
fast for large $E$. An interesting consequence is the appearance of `interrupted
aging' effects for low temperatures, with a correlation function described by Eq.
(\ref{Pi1},\ref{Pi2}). Note that this correlation nearly scales as $t \over t_w$, but
with a systematic bias similar to the one discussed in \cite{BVH}.

The above model is however oversimplified since:

-- no {\it spatial structure} is included and 

-- the particles are independent. 

The next two sections aim at discussing, at least partially, these aspects.

\section {Model of traps in Euclidean space}

\subsection{Continuous-time random walks}

We now consider that the traps live on a d-dimensional hypercubic lattice.
The particle performs a usual random walk on this lattice, but has to wait
for a certain trapping-time $\tau$ before each jump \cite{BG90}.
For a trap of depth $E$, the distribution of the trapping-time $\tau$ 
reads
\be
\Psi_E(\tau)= \Gamma_0 \ e^{-\beta E} \
 e^{ -\left(\Gamma_0 \ e^{-\beta E} \right)\tau}
\ee
If $\rho(E)$ denotes the probability distribution of the depth $E$ of the traps,
 the particle performs a random walk among traps 
with a distribution of trapping times 
\be
\Psi(\tau)= \int_0^{\infty} dE \ \rho(E) \ \Psi_E(\tau)
\label{trapdistr}
\ee
Note that we consider here a model with annealed disorder, where the energy $E$
for a given site changes at each visit of the particle. 
This is justified for dimensions $d > 2$, where the particle rarely visits twice the
same site. The model that we consider is thus the well known `continuous time random
walk' (CTRW) first introduced by Montroll and Scher, and discussed many times \cite{CTRW}.
However, the non stationary properties of the low temperature phase (when it exists)
has, to our knowledge, never been considered -- only the case where the waiting time
$t_w$ is zero has been investigated. We shall thus focus on the following natural
correlation function:  
\be C(\vec q, t+t_w,t_w) = < e^{i \vec q \cdot \big( \vec r (t+t_w)-
\vec r (t_w) \big)} > \ee 
in particular in the case where there is no equilibrium
distribution for (\ref{MEtrap}).

\subsection{Correlation function $C(\vec q, t,t_w=0)$} 

We shall first recall the result for a vanishing waiting time $t_w=0$, which is the 
case
usually considered \cite{CTRW}. One has by definition:
\be
C(\vec q, t,0) = < e^{i \vec q \cdot\big( \vec r (t)- \vec r (0) \big)} >
= \sum_{\vec r} \ e^{i \vec q \cdot \vec r } \ {\cal P} (\vec r, t \vert \vec 0,0)
\ee
where the summation is over all sites of the lattice. The probability density
may be decomposed according to the number of jumps $N$ as
\be
 {\cal P} (\vec r, t \vert \vec 0,0) =\sum_{N=0}^{\infty} Q_N(\vec r) R_t(N)
\ee
where $Q_N(\vec r)$ denotes the probability to be at site $\vec r$ 
 after $N$ jumps for a usual random walk on the lattice. $R_t(N)$ denotes
the probability to perform $N$ jumps in time $t$ and reads in terms of
the distribution $\Psi(\tau)$ of trapping time and Heaviside function $\theta$
\be
R_t(N)= \left(\prod_{i=1}^{N+1} d\tau_i \Psi(\tau_i) \right) 
\theta \left(t-\sum_{i=1}^{N}\tau_i \right)
 \theta \left(\sum_{i=1}^{N+1}\tau_i -t \right) 
\ee
or more explicitly in terms of the probability density $\Psi_N(\tau)$
of the sum $\tau=\sum_{i=1}^{N}\tau_i $ of $N$ independent trapping times 
\be
R_t(N)= \int_0^t d\tau \bigg( \Psi_N(\tau) - \Psi_{N+1}(\tau) \bigg)
\ee
The correlation function may then also be decomposed 
according to the number of jumps $N$ as
\be
C(\vec q, t,0) = \sum_{N=0}^{\infty}  R_t(N) {\hat Q}_N(\vec q)
\ee
where ${\hat Q}_N(\vec q)$ denotes the Fourier transform of a usual random walk
on a d-dimensional hypercubic lattice of lattice spacing $a$
starting from site $\vec 0$  
\be
{\hat Q}_N(\vec q) = \sum_{\vec r} \ e^{i \vec q . \vec r (t)} \ Q_N(\vec r)
= \bigg( {1 \over d} \sum_{\mu=1}^d \cos(q_{\mu}a) \bigg)^N
\ee
It is convenient to introduce the Laplace transform
\be
{\tilde C}(\vec q, \lambda ,0) =\int_0^{\infty} dt \ e^{- \lambda t}  C(\vec q, t,0)
=\sum_{N=0}^{\infty}   {\hat Q}_N(\vec q) {\tilde R}_{\lambda}(N)
\ee
where 
\be
{\tilde R}_{\lambda}(N)=\int_0^{\infty} dt \ e^{- \lambda t} R_t(N)
= {1 \over \lambda} \bigg( {\tilde\Psi}(\lambda) \bigg)^N
\bigg[1- {\tilde\Psi}(\lambda) \bigg]
\ee
in terms of the Laplace transform ${\tilde\Psi}(\lambda)$ of the distribution
 $\Psi(\tau)$ of trapping time
\be
{\tilde\Psi}(\lambda)=\int_0^{\infty} d\tau \ e^{- \lambda \tau} \ \Psi(\tau)
\ee
We obtain finally
\be
{\tilde C}(\vec q, \lambda ,0) = {1 \over \lambda}
 { {1-{\tilde\Psi}(\lambda)Ê} \over {1-{\tilde\Psi}(\lambda)
\bigg(\displaystyle  {1 \over d} \sum_{\mu=1}^d \cos(q_{\mu}a) \bigg)  }}
\label{Correl0}
\ee

\subsection{Correlation function $C(\vec q, t+t_w,t_w) $} 

We now turn to the case of arbitrary waiting time $t_w$ where 
 the correlation function reads now:
\be
C(\vec q, t+t_w,t_w) = \sum_{\vec r} \ \sum_{\vec r_w} \ 
 e^{i \vec q \cdot ( \vec r -\vec r_w) } \ \int_0^{\infty} dE_w \ 
{\cal P} (\vec r, t+ t_w \vert  \vec r_w, E_w, t_w)
{\cal P} (\vec r_w, E_w, t_w \vert \vec 0,0)
 \ee
It is convenient to reorganize this expression as follows
\be
C(\vec q, t+t_w,t_w) =
\int_0^{\infty} dE_w \ \sum_{\vec r_w} \ {\cal P} (\vec r_w, E_w, t_w \vert \vec 0,0)
\sum_{\vec r} \  e^{i \vec q \cdot ( \vec r -\vec r_w) } \
{\cal P} (\vec r, t+ t_w \vert  \vec r_w, E_w, t_w)
\ee 
and to decompose the conditional probability 
${\cal P} (\vec r, t+ t_w \vert \vec  r_w, E_w, t_w)$ into
\be
{\cal P} (\vec r, t+ t_w \vert \vec  r_w, E_w, t_w)=
\delta_{\vec r, \vec r_w} \ e^{ \displaystyle
-\left(\Gamma_0 e^{- \beta E_w}\right)t } + \int_0^t d\tau \ \Psi_{E_w} (\tau) 
{1 \over {2d}} \ 
\sum_{\vec e} 
{\cal P} (\vec r, t-\tau \vert \vec r_w + \vec e ,0)
\ee
The first term takes into account the probability to stay in the trap $\vec r_w$
of energy $E_w$ during the time interval $[t_w, t_w+t]$.
The second term describes the probability to jump at the time $t_w+t$
out of the trap of energy $E_w$ to go to one of the $2d$ neighbours $\vec r_w + \vec e$ 
of $ \vec r_w$, and where the `aging' process starts anew.
One has therefore
\be
\sum_{\vec r} \  e^{i \vec \cdot ( \vec r -\vec r_w) } \
{\cal P} (\vec r, t+ t_w \vert \vec r_w, E_w, t_w)
= e^{ \displaystyle -\left(\Gamma_0 e^{- \beta E_w}\right) t}
\ee
\be 
+\int_0^t d\tau \Psi_{E_w} (\tau)
\bigg( {1 \over d} \sum_{\mu=1}^d \cos(q_{\mu}a) \bigg)
C(\vec q, t-\tau,t_w=0)
\ee
Noticing that
\be 
\sum_{\vec r_w} \ {\cal P} (\vec r_w, E_w, t_w \vert \vec 0,0) =P(E_w,t_w)
\ee
is simply the probability distribution studied in the previous section (\ref{MEtrap}),
and that
\be
\int_0^{\infty} dE_w \ P(E_w,t_w) e^{ \displaystyle - \left(\Gamma_0 e^{- \beta E_w}\right)t }
=\Pi(t+t_w,t_w)
\ee
is the correlation function (\ref{Pidef}) introduced previously,
we obtain finally
\be
C(\vec q, t+t_w,t_w) = \Pi(t+t_w,t_w) - 
\bigg( {1 \over d} \sum_{\mu=1}^d \cos(q_{\mu}a) \bigg)
\int_0^t d\tau \ C(\vec q, t-\tau,t_w=0) \
 { \partial \over {\partial \tau} } \Pi(\tau +t_w,t_w)
\ee

This convolution product leads to introduce the Laplace transforms
\be
{\tilde C}(\vec q, \lambda ,t_w) =
\int_0^{\infty} dt \ e^{- \lambda t}  \ C(\vec q, t+t_w,t_w)
\ee
and
\be
{\tilde \Pi}( \lambda ,t_w) =
\int_0^{\infty} dt \ e^{- \lambda t} \ \Pi( t+t_w,t_w)
\ee
to get the simpler relation
\be
{\tilde C}(\vec q, \lambda ,t_w) =
{\tilde \Pi}( \lambda ,t_w) \ \bigg[1-
\lambda \bigg( {1 \over d} \sum_{\mu=1}^d \cos(q_{\mu}a) \bigg)
{\tilde C}(\vec q, \lambda ,0) \bigg]
+ \bigg( {1 \over d} \sum_{\mu=1}^d \cos(q_{\mu}a) \bigg)
{\tilde C}(\vec q, \lambda ,0) 
\ee
Equation (\ref{Correl0}) gives finally
\be
{\tilde C}(\vec q, \lambda ,t_w) = {\tilde \Pi}( \lambda ,t_w) \
\left[ { {1-\bigg(\displaystyle  {1 \over d} \sum_{\mu=1}^d \cos(q_{\mu}a) \bigg)}
 \over {1-{\tilde \Psi}(\lambda)
 \bigg( \displaystyle {1 \over d} \sum_{\mu=1}^d \cos(q_{\mu}a) \bigg)}}  \right]
+{1 \over \lambda} { {1-{\tilde \Psi}(\lambda)} \over {
\bigg(\displaystyle  {1 \over d} \sum_{\mu=1}^d \cos(q_{\mu}a) \bigg)^{-1}
-{\tilde \Psi}(\lambda)}}
\label{Correlw}
\ee
 
Another quantity of interest in glasses is the susceptibility, 
which is related to the above correlation function through:
 
\be
\chi(\vec q,\omega,t_w) = 1 + i\omega \int_0^\infty dt \  e^{i \omega t} 
C(\vec q,t_w+t,t_w) \equiv 1 + i \omega \ \tilde C(\vec q, -i \omega ,t_w)\label{chi}
\ee

(see however \cite{CuKu} for a discussion of the Fluctuation-Dissipation theorem in this context).

\subsection{Case of exponential density of states at low temperature}

For the exponential density of states $\rho(E)= \beta_0 \exp(-\beta_0 E)$,
there is no equilibrium distribution for the process (\ref{MEtrap})
when $x \equiv {T \over T_0} <1$. The correlation function $\Pi$
is known in this case (\ref{Piexp}) and its Laplace transform reads 
\be
{\tilde \Pi}( \lambda ,t_w) \simeq  {1 \over t_w} \int_0^{\infty} du \
 e^{ \displaystyle - (\lambda t_w) u} \ \  { {\sin \pi x} \over {\pi }}
\int_{ { u } \over {1+uÊ}}^{1} dv 
\left(1-v \right)^{x -1} v^{- x} 
\label{PiexpL}
\ee
with the asymptotic behaviors
\be
{\tilde \Pi}( \lambda ,t_w) \opsimeq_{\lambda t_w \ll 1}
{1 \over {\Gamma(1+x)}} {Ê{(\lambda t_w)^{x}} \over {\lambda}}
\ee
\be
{\tilde \Pi}( \lambda ,t_w) \opsimeq_{\lambda t_w \gg 1}
{1 \over {\lambda}} \bigg[1-{1 \over {\Gamma(x)}} {1 \over {
(\lambda t_w)^{1-x}}} \bigg]
\ee
 The distribution of trapping times 
\be
\Psi(\tau)= \int_0^{\infty} dE \ \rho(E) \ \Gamma_0 \ e^{-\beta E} \
 e^{- \left(\Gamma_0 \ e^{-\beta E} \right)\tau}
={ {x \Gamma_0} \over {\left( \Gamma_0 \tau \right)^{1+x} }}
\int_0^{ \Gamma_0 \tau} dy \ y^{x} e^{-y}
\ee
presents the slow algebraic decay
\be
\Psi(\tau) \opsimeq_{(\Gamma_0 \tau) \to \infty}
{ {x  \Gamma(1+x)} \over
 { \Gamma_0^x \tau^{1+x} }} 
\ee
and its Laplace transform presents therefore the non-analytic behavior
\be
{\tilde\Psi}(\lambda)=\int_0^{\infty} d\tau \ e^{- \lambda \tau} \ \Psi(\tau)
\opsimeq_{ \left({\lambda \over \Gamma_0}\right) \to 0^+}
 1-\ {1 \over \sinc \pi x}
\left({\lambda \over \Gamma_0}\right)^{x} \qquad
\label{psila}
\ee
Moreover, one has for wavevectors such that $ {1 \over q}$ is
much larger than the lattice spacing $a$
 \be
\bigg( {1 \over d} \sum_{\mu=1}^d \cos(q_{\mu}a) \bigg) \opsimeq_{\vert q \vert \ll 
{1 \over a}} 1- {a^2 \over {2d}} {\vec q}^2 +...
\label{sumcos}
\ee
It is convenient to introduce the time scale $t_q$ 
\be
 (\Gamma_0 t_q)^{x}= {2d \over (qa)^2}  
\ee 
corresponding to the typical time needed by the particle to spread over a region of size $1/q$. There are therefore three time scales in (\ref{Correlw}) : $\lambda^{-1}, t_q,t_w$.
We are interested in the region where all three are much bigger than the
microscopic time scale $\Gamma_0^{-1}$ : $ \lambda^{-1} \gg {1 \over \Gamma_0}$ 
(\ref{PiexpL}-\ref{psila}) ,
$t_q \gg  {1 \over \Gamma_0}$ (\ref{sumcos}) and $t_w \gg  {1 \over \Gamma_0}$
(\ref{PiexpL}).
Still we have to distinguish various regimes in (\ref{Correlw}).
We give the asymptotic results for the correlation function $C(\vec q, t+t_w,t_w)$ 
after the inversion of the Laplace transform (\ref{Correlw}):
\be
C(\vec q, t+t_w,t_w) \simeq   
 1-    {\sinc [\pi (1-x)]}  
\left({t \over t_w}\right)^{1-x } \qquad \hbox{for} \ t_q \ll t \ll t_w  
\ee    
\be
C(\vec q, t+t_w,t_w) \simeq   {\sinc [\pi x]} 
\left({t \over t_w}\right)^{-x }   
\qquad \hbox{for} \  t_q \ll t_w \ll t   \ 
\ee

%For $t \ll t_w, t_q$,
\be
C(\vec q, t+t_w,t_w) \simeq 1- {\sinc [\pi x] \over {\Gamma(x)}} 
{t \over {t_q^{x} t_w^{1-x}}}
\qquad \hbox{for} \   t \ll t_w  \ \ \hbox{and}\  t \ll  t_q   \ 
\ee

%For $t_w \ll t \ll t_q$

\be
C(\vec q, t+t_w,t_w) \simeq 1-{\sinc [\pi x]\over { \Gamma(1+x)}}
\left({t \over t_q} \right)^{x}
\qquad \hbox{for} \ t_w \ll t \ll t_q
\ee 

%For $t_w \ll t_q \ll t$
\be
C(\vec q, t+t_w,t_w) \simeq \Gamma(1+x)
\left({t \over t_q}\right)^{-x }
\qquad \hbox{for} \ t_w \ll t_q \ll t 
\ee

There are three interesting points to notice:

$\bullet$ For $qa$ very large, such that $t_q \ll t,t_w$, we find the same asymptotic behaviors
 as for $\Pi(t+t_w,t_w )$ (\ref{agingexp0} -\ref{agingexpinf}) in the previous
section. Physically, this means that as soon as the particle has jumped once, the
rapidly oscillating correlation function averages to zero. Hence, only the particles
which have not yet moved contribute to the correlation.

$\bullet$ There are two regimes where the correlation function behaves similarly to a
 stretched exponential at small times, when $t_q \ll t \ll t_w$ or $t_w \ll t \ll
t_q$.  The exponent $\alpha$ of this stretched exponential is however different in
both cases: it is equal to $\alpha=x$ when $t_w \ll t \ll t_q$, and equal to
$\alpha=1-x$ in the other case.

One should however stress once more that the present trap model is unable to 
explain why there should be a relation between the exponent $\alpha$ describing
the decay of the correlation in the $\alpha$ regime which we discuss here, and the 
shape of the relaxation in the short-time ($\beta$) regime, which corresponds to
intra-trap dynamics. That such a link exists is one of the major prediction of the
MCT, which should also exist deep in the glass \cite{us}
phase, where aging effects similar to those discussed here are present.

$\bullet$ The frequency dependent susceptibility defined by Eq. (\ref{chi}) behaves
{\it in a Cole-Cole fashion}: \be 
\chi(\vec q,\omega,t_w=0) \simeq {1 \over 1+ (-i\omega t_q)^x}
\ee
for $\omega \ll \Gamma_0$, $qa \ll 1$. This has been 
emphasized in e.g. \cite{Japrl}. However, in the aging regime $\omega t_w \gg 1$, the
behaviour of $\chi$ is given by: 
\be
\chi(\vec q,\omega,t_w) \simeq { 1\over \Gamma(x) (-i \omega t_w)^{1-x} }
\ee
A similar expression was obtained in the context of spin-glasses in \cite{BD95}. 

\section {Model of interacting particles in traps}

\subsection {Definition of the model}

In the above sections, we have considered models for which the motion of a given 
particle out of its trap does not affect the potential seen by the others. In order to
model this effect, we have proposed in \cite{BCM95} to add to model (\ref{MEtrap}) a
diffusion term in energy space proportional to the mean hopping rate itself 
$\omega(t)=\int_0^{\infty} dE  \ e^{-\beta E} \ \P$. The equation for $P(E,t)$ then
reads:

\be
{1 \over \Gamma_0} {\partial P \over \partial t} = - e^{-\beta E} \ P
 + \omega(t) \rho(E) +
\omega(t) D {\partial \over \partial E} \left[\rho(E)  {\partial P \over
\partial E} - P {d \rho \over d E}\right] 
\label{MEinter}
\ee

This diffusion term  expresses the fact that every `hop' induces a small change in all the neighbouring $\epsilon$'s. Assuming that the transition rate is proportional to the final density of states, the contribution of such an effect to the master equation
 a priori reads : $\omega(t) \int 
d\epsilon' \ {\cal T}(\vert \epsilon-\epsilon'\vert) \{P(\epsilon',t)\rho(\epsilon) 
 - \P \rho(\epsilon')\}$.
In the limit where the width of $ {\cal T}(\vert \epsilon-\epsilon'\vert)$ is small, 
justified in a
mean-field limit where the number of neighbours is large, 
this term reduces to the diffusion like term in (\ref{MEinter}), with an effective 
diffusion constant $D$ 
proportional to the width of $\cal T$.  More general forms for this diffusion term  
will be discussed below. 

As before, equation (\ref{MEinter}) has to be supplemented by some initial condition 
$P(E,t=0)=P_0(E)$
and by a `hard wall' boundary condition at $E=0$
\be
\left[\rho(E)  {\partial \P \over
\partial E} - \P {d \rho \over d E}\right] \bigg\vert_{E=0} =0
\label{bc}
\ee
to ensure the conservation of probability.

\subsection {Existence of a stationary distribution}

There exists a stationary distribution $\Pe$ at temperature $T=\beta^{-1}$ only if
the equation
\be
-D { {d^2 \Pe} \over dE^2Ê} + \left[ D {\rho''(E) \over \rho(E)} 
+ { {e^{- \beta E}} \over {\omega_{eq} \rho(E) }} \right] \Pe =1
\label{eqPeq}
\ee
supplemented by the boundary condition (\ref{bc}) admits a normalizable solution.
The discussion again depends on $T_0=\beta_0^{-1}$ introduced in (\ref{defbeta0}). 
\hfill \break
$\bullet$ For $T>T_0 $,
the asymptotic behavior at high energy of the solution of (\ref{eqPeq}) reads
\be
\Pe \opsimeq_{E \to \infty}\omega_{eq} e^{\beta E} \rho(E)
\label{asy}
\ee
which is normalizable since $\beta < \beta_0$.\hfill \break
$\bullet$ For $T<T_0$, equation (\ref{eqPeq}) can be approximated
 at high energy by 
\be
 { {d^2 \Pe} \over dE^2Ê} -  {\rho''(E) \over \rho(E)}  \Pe = - {1 \over D}
\label{eqPeqs}
\ee
The general solution reads using arbitrary constants $A$ and $B$
\be
A g_1(E) +B g_2(E) 
 + {1 \over D} \left[g_1(E) \int_0^E dv\  g_2(v) +g_2(E)  \int_E^{\infty} dv \ g_1(v) \right]
\label{eqPeqsol}
\ee
 in terms of the two independent solutions
\be
g_1(E)=\rho(E) \qquad \hbox{and} \qquad g_2(E)=\rho(E) \int_0^{E}{d u \over \rho^2(u) } 
\label{homo}
\ee
of the homogeneous equation.
Taking into account that $\rho(E)$ decays at least exponentially
at large $E$ for $\beta_0 < + \infty$, one may show that there is no 
normalisable solution.

In summary, the conditions for the existence of a stationary distribution in term
of $T_0$ are therefore exactly the same as in the case $D=0$ (\ref{defbeta0}).

The case of an exponential density of states was considered in \cite{BCM95}. 
For completeness, we give the explicit form of the equilibrium distribution:

\be
\Pe= {\cal N} \left[ K_\nu(z) {I_{\nu-1}(z_0) \over K_{\nu-1}(z_0)} 
{\cal K}_\nu(z_0) + K_\nu(z) ({\cal I}_\nu(z_0) - {\cal I}_\nu(z)) + I_\nu(z)
{\cal K}_\nu(z)\right]
\ee
where $\nu= {2 T \over T-T_0}$, $z \equiv z_0 \exp {E \over \nu T_0}$, 
and $z_0 = {\nu T_0^{3/2}\over \sqrt{ D}}{\sqrt{\Gamma_0} \over \sqrt{\Gamma}}$.
$I_\nu$ and $K_\nu$ are the Bessel functions of order $\nu$, and ${\cal K}_\nu(x)
\equiv \int_x^\infty {du \over u} K_\nu(u)$, ${\cal I}_\nu(z) \equiv \int_0^z {du
\over u} I_\nu(u)$. $\cal N$ and $\Gamma$ are fixed by the normalisation of $\Pe$ and
the boundary
 condition
(\ref{bc}), which lead to the following equation:

\be
{D \over \nu^3 T_0^3} =  {I_{\nu-1}(z_0) \over K_{\nu-1}(z_0)} 
[{\cal K}_\nu(z_0)]^2 + 2 \int_{z_0}^\infty {du \over u} I_\nu(u) {\cal K}_\nu(u) 
\ee

\subsection {Aging in the low temperature phase}

When there is no equilibrium distribution (\ref{omeq}), we may proceed as in the case
 $D=0$
 (section \ref{sec-aging}) and look for a scaling
solution of the form (\ref{scaling}) for the equation (\ref{MEinter}).
 The resulting equation for the dimensionless function $\phi$
again only admits some non trivial limit as $t \to \infty$ 
 if the left-handside of (\ref{eqphi}) does.
For example, in the case of an exponential density of states
$\rho(E) = \beta_0 \ e^{- \beta_0 E}$, in which there is no equilibrium distribution
when $x \equiv {T \over T_0} \in ] 0,1[$,
the function $\phi$ must present the singularity (\ref{sing0})
\be
\phi(u) \opsimeq_{u \to 0} \gamma \ u^{-x}
\label{sing0bis}
\ee
 The equation for $\phi$ now generalizes
equation (\ref{eqphiexp}) 
\be
\Delta u^3 {d^2 \phi \over du^2} + \big[ 3\Delta u^2+ u^{2+x} \big] {d \phi \over du} +
 \big[\Delta(1-x^3)u+ (u-1) u^{x} \big] \phi(u) = -\gamma 
\label{eqphiexpD}
\ee
where $\Delta = \gamma D \beta^3$. The normalisation constant $\gamma$
is determined by the normalization $\int_0^{\infty} du \ \phi(u)=1$.
Contrary to the case $D=0$ (\ref{solphi}), the solution $\phi$ 
cannot be explicitly written. However, its asymptotic behavior at large $u$
is easily obtained from (\ref{eqphiexpD})
\be
\phi(u) \opsimeq_{u \to \infty} {\gamma \over x} \  u^{-1-x}
\label{singinftybis}
\ee

The aging behaviour (\ref{Piaging}) of the correlation function (\ref{Pidef})
still holds as a consequence of the scaling form (\ref{scaling}).
The asymptotic behaviors (\ref{sing0bis}-\ref{singinftybis})
of the scaling function $\phi$ 
induce the asymptotic expressions generalizing (\ref{agingexp0} - \ref{agingexpinf})
 for the correlation function
\be
\Pi(t_w,t_w+t ) \simeq   
 1-  \gamma \ { {\Gamma (x)} \over { (1-x) }} 
\left({t \over t_w}\right)^{1-x } \qquad \hbox{for} \ t \ll t_w  
\label{agingexpasyD0}
\ee    
\be
\Pi(t_w,t_w+t ) \simeq  \gamma \ { {\Gamma(x)} \over {x }}
\left({t \over t_w}\right)^{-x }   \qquad \hbox{for} \ t \gg t_w   \ 
 \label{agingexpasyDinf}
\ee
The presence of the diffusion term in (\ref{MEinter}) therefore only affects
the normalisation constant $\gamma$, but does not change the asymptotic time dependence 
of the correlation function $\Pi(t_w,t_w+t )$ of the model (\ref{MEtrap}).
This suggests that the 
difference between the model considered in section (\ref{sec-aging}) 
(corresponding to $D \equiv 0$) and the `annealed' model 
considered here is, to some extent, irrelevant.

\subsection{Possible generalizations of model (\ref{MEinter})   }

The model (\ref{MEinter}) may in fact be generalized through the introduction of 
an energy dependent diffusion constant, i.e.:
\be
{1 \over \Gamma_0} {\partial P \over \partial t} = - e^{-\beta E} \ P
 + \omega(t) \rho(E) +
\omega(t) {\partial \over \partial E} \bigg(  D(E) \left[\rho(E)  {\partial P \over
\partial E} - P {d \rho \over d E}\right]  \bigg)
\ee
with the same boundary condition (\ref{bc}) as before
to ensure the conservation of probability. $\rho(E)$ still corresponds to the
 equilibrium 
distribution at infinite temperature ($\beta=0$). The choice 
$D(E) = {D_0 \over \rho(E)}$, for example, would correspond to an `entropy' biased
diffusion in energy space, with a driving force proportional to ${d \log \rho \over d
E}$, and an effective diffusion constant independent of $E$.

A careful study of the existence of a stationary solution to this equation 
shows that there are two cases. \hfill \break
$\bullet$ If $  \displaystyle \lim_{E \to \infty} {1 \over E} \log D(E) \leq 0 $,
the transition takes place as before at the inverse temperature $\beta_0$  
(\ref{defbeta0}). 
\hfill \break
$\bullet$ If $  \displaystyle \lim_{E \to \infty} {1 \over E} \log D(E) > 0 $, there
exists a stationary distribution at any finite temperature. In the case where 
$D(E) = {D_0 \over \rho(E)}$ and $\rho(E)$ exponential, for example, one finds that the large $E$ behaviour of $\Pe$ is of the form:
\be
\Pe \opsimeq_{E \to \infty} A e^{-\beta_0 E} + B e^{-2 \beta_0 E}
\ee
showing that the correlation function now decays as $C_{\rm{eq}}(t) \propto 
t^{-\left(1 + {T \over T_0}\right)}$ for large times. Note that the terminal time scale
$\tau_1 = \int_0^\infty dt \ C_{\rm{eq}}(t)$ diverges when $T < T_0$, although
no asymptotic aging effects appear in this temperature regime.

\vskip 2 true cm 
\section{Conclusion}

We have studied in this paper a model of particle hopping between energy `traps'
with an arbitrary density of energy barriers $\rho(E)$. As emphasized in
\cite{Bou92,BD95,BCM95}, the case where  $\rho(E)$  decays exponentially is special
because it leads to a true dynamical  phase transition between a high temperature
phase and a low temperature aging phase.  More generally, however, on expects that for
a large class of $\rho(E)$, `interrupted' aging effects appear at low enough
temperatures, with an ergodic time  growing faster than exponentially. It
would be interesting to look systematically for aging effects experimentally \cite{Struik}. Furthermore, the
relaxation functions have a strongly stretched shape (see e.g. Fig. 1-b), which can be
fitted as stretched  exponentials. The case where the traps are organized on a $d$
dimensional lattice is slightly more involved, since a new time scale $t_q$ appears,
which is a wave-vector dependent relaxation time. A schematic way of  modelling the
{\it interactions} between the particles, reflecting the fact that when one particle
moves, the potential energy seen by its neighbours changes, was investigated in
section 4. The conclusion regarding the existence of a dynamical transition was found
to depend on the shape of the effective diffusion constant $D(E)$. In the case where
the transition survives, the role of this `interaction' is irrelevant.

In conclusion, many of the observed features of glassy systems can be accounted for 
within a picture of independent particles trapped in random potential wells, without
any obvious collective effects \cite{Bass,Vilgis,Odagaki1,Odagaki2,BCM95}. 
How deep is the link between this picture and the
Mode-Coupling Theory is not yet clear. However, the MCT relies on the existence of a
true critical point $T_c$ , with two important consequences which are beyond the grasp
of `trap' models: the strong link between the $\alpha$ and $\beta$ regime and the
critical behaviour of the plateau value in $C_{\rm{eq}}(t)$ as $\sqrt{T_c - T}$
\cite{Gotze}. These predictions are however hard to test since the critical point is
supposed to be blurred by `activated processes'. As suggested in \cite{us}, a crucial
test of MCT could be performed by working deep below $T_c$, where `activated
processes' are frozen but where strong aging effects should appear.

\vskip 1 true cm 

\centerline{\bf Appendix A } 
\centerline{}
\centerline{\bf Relaxation towards the equilibrium 
distribution in model (\ref{MEinter})}

\vskip 0.5 true cm

In the case $D>0$, equation (\ref{MEinter}) is nonlinear and cannot be solved exactly.
However, we may introduce some approximations to study the relaxation at large time.

\vskip 0.5 true cm

\leftline{ \bf Approximate description of the relaxation}
\vskip 0.5 true cm

To study the relaxation towards the equilibrium distribution $\Pe$, we may set
\be
P(E,t) = \Pe + \eta(E,t)
\label{eta}
\ee
into (\ref{MEinter}), linearize the corresponding equation in the perturbation 
$\eta$, and replace $\Pe$
by the expression $\Pe \simeq \om e^{+\beta E} \ \rho(E)$
 which happens to be a very good approximation except
 in the vicinity of $E=0$. We obtain finally 
\ba
\nonumber {1 \over \Gamma_0} {\partial \eta \over \partial t} = - e^{-\beta E} \ \eta (E,t)
 +  \rho(E)  \int_0^{\infty} du \  e^{- \beta u} \  \eta(u,t) \\
+ \om D  \left[\rho(E)  {\partial^2 \eta \over
\partial^2 E} - \eta { \rho''(E) }\right] 
\label{eqeta}
\ea

We may look for the solution through its decomposition onto
a relaxation spectrum
\be
\eta(E,t)  = \int_0^{\infty} d\lambda \ f_{\lambda}(E) \ e^{- \lambda \Gamma_0 t}  
\label{Relaxspdec}
\ee
with the condition 
\be
\int_0^{\infty} dE \ f_{\lambda}(E) =0
\label{normfsp}
\ee
 to insure the normalization of the probability density $\P$ (\ref{eta}).
The equation for $f_{\lambda}$ (\ref{eqeta})
\be
-D \om{  {d^2 f_{\lambda}} \over {d^2 E} } + \left[D \om {\rho''(E) \over \rho(E)}
+ { { e^{-\beta E} -\lambda} \over {\rho(E)}} \right] f_{\lambda}(E) = C_{\lambda}
\label{eqflambda}
\ee
where 
\be
C_{\lambda} = \int_0^{\infty} d\lambda \ f_{\lambda}(E) \ e^{- \beta E}
\label{Clambda}
\ee
has to be supplemented by the boundary condition (\ref{bc})
\be
\left[\rho(E)  {d f_{\lambda} \over
d E} - f_{\lambda} {d \rho \over d E}\right] \bigg\vert_{E=0} =0
\label{bcf}
\ee
In fact, the self-consistency condition (\ref{Clambda}) is automatically satisfied
once $f_{\lambda}$ is solution of (\ref{eqflambda}-\ref{normfsp}-\ref{bcf}).

Equations (\ref{eqflambda}-\ref{normfsp}-\ref{bcf}) imply that
\be 
\left( \lambda - \lambda' \right) \int_0^{\infty} {dE  \over \rho(E)}
 f_{\lambda}(E) f_{\lambda'}(E) =0
\label{ortho}
\ee
The functions $ f_{\lambda}(E)$ including $\Pe =  f_{0}(E)$ 
are therefore orthogonal to one another with respect to the measure ${dE  \over \rho(E)}$.
One may choose to orthonormalize this set of functions according to the scalar product
\be 
 \int_0^{\infty} {dE  \over \rho(E)} \ 
 f_{\lambda}(E) f_{\lambda'}(E) = \delta(\lambda-\lambda')
\label{orthon}
\ee
The development of $P(E,t)$ onto this basis then reads
\be
P(E,t) = \Pe+ \int_0^{\infty} d\lambda \ a(\lambda) \  f_{\lambda}(E) \ e^{- \lambda \Gamma_0 t}  
\label{devbasis}
\ee
where the coefficients $a(\lambda)$ are simply obtained through scalar product 
with the initial condition $P(E,t=0)$
\be 
a(\lambda) = \int_0^{\infty} {dE  \over \rho(E)} \ 
 f_{\lambda}(E) P(E,0) 
\label{coefs}
\ee

\vskip 0.5 true cm

\leftline{\bf Nature of the relation spectrum}

\vskip 0.5 true cm

To determine the relaxation spectrum, it is convenient to transform
(\ref{orthon}) into a usual scalar product
\be
\delta(\lambda-\lambda') =
 \int_0^{\infty} {dE  \over \rho(E)} \  f_{\lambda}(E) f_{\lambda'}(E) = 
\int_{u_0}^{\infty} du \  \Psi_{\lambda}(u) \Psi_{\lambda'}(u)
\label{usorthon}
\ee
 through the change of variable
\be
E \longrightarrow u(E) = \int_{u_0}^{E} {d E' \over \sqrt{\rho(E')} }
\label{chvar}
\ee
or conversely 
\be
E(u) = \int_0^{u} {dv \over h(v)} \qquad \hbox{where} \qquad  h(u)
 = {1 \over \sqrt{\rho[E(u)]} }
\label{reci}
\ee
 and a change of functions
\be
 f_{\lambda}(E) \longrightarrow \Psi_{\lambda}(u) =
 { f_{\lambda}[E(u)] \over {\left(\rho[E(u)]\right)^{1 \over 4} } }
\label{chfonc}
\ee
The equation for the new function $\Psi_{\lambda}(u)$ (\ref{eqflambda}) 
\be
-{d^2 \Psi_{\lambda} \over d^2 u} + \left[ V(u)
+ { {e^{-\beta E(u)} -\lambda} \over {D \om}} \right] \Psi_{\lambda}(u) = 
{ { C_{\lambda}} \over
  { D \om \left[h(u)\right]^{3 \over 2} } }
\label{eqpsilambda}
\ee
where 
\be
V(u) = {15 \over 4} \left({h'(u) \over h(u) }\right)^2 -{3 \over 2} {h''(u) \over h(u)}
\label{Vu}
\ee
  has to be supplemented by the boundary
condition at $u_0=u(E=0)$
\be
 \left[ {d \Psi_{\lambda} \over du} + 
{3 \over 2} {h'(u) \over h(u) } {\Psi_{\lambda} } \right] \bigg\vert_{u=u_0} =0
\label{bcpsi}
\ee
and by the condition (\ref{normfsp})
\be
\int_{u_0}^{\infty}  du \ {\Psi_{\lambda}(u) \over [h(u)]^{3 \over 2} } =0
\label{normpsi}
\ee
To discuss the nature of spectrum, the only important property of the potential $V(u)$
is that
\be
V(u)  \operarrow_{u \to \infty} 0
\label{Vinf}
\ee
This comes from the fact that
 $\rho(E)$ presents a rapid enough decay as $E \to \infty$ to be integrable.   

At infinite temperature ($\beta=0$), the constant $C_{\lambda}$ (\ref{Clambda})
on the left-hand side of (\ref{eqpsilambda}) vanishes (\ref{normfsp}). 
The asymptotic form of equation (\ref{eqpsilambda}) as $u \to \infty$
\be
  {d^2 \Psi_{\lambda} \over d^2 u} = - \left( { {\lambda -1} \over {D \om}} \right)   \Psi_{\lambda}(u) 
\label{eqpsiasy0}
\ee
admits two independent oscillatory solutions for $\lambda > 1$,
and the boundary condition at $u_0$ determines
the suitable linear combination to be takken for $\Psi_{\lambda}$
up to a normalization constant.
However, for $\lambda < 1$, 
 only the exponentially damped solution is acceptable at infinity and
the boundary condition at $u_0$ cannot be satisfied.
So we finally obtain that the relaxation spectrum at infinite temperature ($\beta=0$)
consists in a continuum above the gap $\Gamma_0$
\be
P(E,t) = \rho(E) + \int_1^{\infty} d\lambda \ a(\lambda) f_{\lambda}(E) 
\ e^{- \lambda \Gamma_0 t}  
\label{devTinf}
\ee
to be compared with the previous result for the limiting case $D=0$ (\ref{Relaxinf})
 
At finite temperature ($\beta>0$), the asymptotic form of the
homogeneous equation corresponding to (\ref{eqpsilambda}) in the limit $u \to \infty$
\be
{d^2 \chi_{\lambda} \over d^2 u} = - {\lambda \over {D \om}} \     \chi_{\lambda}(u) 
\label{eqpsiasy}
\ee
admits always oscillatory solutions for $\lambda>0$. The relaxation spectrum
 at finite temperature ($\beta>0$)
consists therefore in a continuum starting from $\lambda=0$ (\ref{devbasis}).
This will produce an algebraic relaxation determined by the behavior of the function
$\phi_{\lambda}$ in the limit $\lambda \to 0$.

\vskip 0.5 true cm

\leftline{ \bf Example of the exponential density of states }

\vskip 0.5 true cm

We may apply the previous general theory to the particular case of
the exponential distribution for $\rho(E)$, 
for which there exists a stationary 
distribution in the domain of high-temperature $ x \equiv {\beta_0 \over \beta} >1$.
The change of variables (\ref{chvar})
\be
E \longrightarrow u(E) =   u_0 \ e^{ {\beta_0 \over 2} E}
\qquad \hbox{where} \qquad  u_0 = {2 \over { \left(\beta_0\right)^{3 \over 2}}}
\label{chvarexp}
\ee
and the function $h(u)$ (\ref{reci})  
\be
 h(u) = {\beta_0 \over 2} u
\label{hexp}
\ee
are simple enough to give an algebraic form to the potential involved
in the inhomogeneous Schr\"odinger equation (\ref{eqpsilambda})
satisfied by the new function $\Psi_{\lambda}(u)$     
\be
-{d^2 \Psi_{\lambda} \over d^2 u} + \left[ {15 \over 4} {1 \over u^2}
+ {1 \over {D \om}} \left( \ {u_0 \over u} \right)^{2 \over x}
 - {\lambda \over {D \om}} \right] \Psi_{\lambda}(u) = 
{ C_{\lambda}  \over {D \om}}   \left( {2 \over {\beta_0 u}} \right)^{3 \over 2} 
\label{eqpsilambdaexp}
\ee

At infinite temperature ($\om=1$ ; $x=\infty$),
the function $f_{\lambda}(E)$ involved in the decomposition (\ref{devTinf}) 
reads in terms of Bessel functions $J_{\nu}$ and $Y_{\nu}$
\be
f_{\lambda}(E) =  {\cal N}_k \ 
 \bigg[ Y_1(k u_0)\  J_2 \big[k u(E) \big] - J_1(k u_0) \ Y_2 \big[k u(E) \big] \bigg]
\label{Bessel}
\ee
with $k=\sqrt{ {\lambda -1} \over D}$ and
\be
{\cal N}_k ={1 \over \sqrt{D \beta_0 \Gamma_0 } } {1 \over \sqrt{ J_1^2(ku_0) +Y_1^2(ku_0) } }
\label{normBessel}
\ee
and (\ref{devTinf}) may be written with the notation $\lambda_k =1+ D k^2$ as
\be
P(E,t) = \rho(E) + 2 D e^{-  \Gamma_0 t}
\int_0^{\infty} k \ dk \  a(\lambda_k) \ f_{\lambda_k}(E) 
\ e^{- D \Gamma_0 k^2 t}  
\label{devTinfexp}
\ee
In the limit $k \to 0$, the behaviors $f_{\lambda_k} \propto k^2$
 and  $a(\lambda_k) \propto k^2$ for a generic initial condition (see \ref{coefs}) 
gives the asymptotic algebraic correction to the exponential $e^{-  \Gamma_0 t}$
\be
P(E,t) - \rho(E) \oppropto_{t \to \infty} {e^{-  \Gamma_0 t} \over (\Gamma_0 t)^3 } 
\label{expalge}
\ee 
to be compared with (\ref{Relaxinf}).
The presence of the diffusion term in (\ref{MEinter}) representing the interaction
between particles thus tends to accelerate the relaxation in comparison to
the case (\ref{MEtrap}) of independent particles.

\vskip 2 true cm

\leftline{\bf Acknowledgments}

\vskip 0.5 true cm

\noindent It is a pleasure to thank A. Comtet for his collaboration 
in the previous letter \cite{BCM95} and in the early stage of the present work.
We also want to thank  L. Cugliandolo, D. Dean, J. Kurchan, C. Godr\`eche and M. M\'ezard 
for many important discussions on these
subjects.
 
\vskip 2 true cm

\leftline{\bf Figure Captions}

{\bf Fig 1}  Correlation function $C_{\rm {eq}}$(t)  (multiplied by a `mock' $
\beta$ relaxation $C_\beta(t) \equiv \exp -{q^2 r^2(t)\over 2}$, where $r(t)$ describes a diffusive motion
 in an harmonic potential well: $r^2(t) = \xi_0^2 [1 - \exp (-{t \over 
\tau_0})]$; $\xi_0$ can be thought as the `size' of the cage, 
and ${\xi_0^2 \over \tau_0}$ of the order of the high temperature diffusion constant).

\noindent {\bf Fig 1-a}  Exponential density of states. \hfill \break
Plot of $C_{\rm {eq}}C_\beta(t)$ versus $\log_{10}(\Gamma_0t)$ for $q\xi_0=0.5$, $\tau_0= 5 \Gamma_0^{-1}$, and 
${T \over T_0}=2.,\ 1.1,\ 1.03$. 

\noindent {\bf Fig 1-b} Gaussian density of states. \hfill \break
Plot of $C_{\rm {eq}}C_\beta(t)$ versus $\log_{10}(\Gamma_0t)$ for $q\xi_0=0.5$, $\tau_0= 5 \Gamma_0^{-1}$, and 
${T \over E_c}=0.5,\ 0.25,\ 0.15$. Note that the plateau observed for the lowest temperature eventually decays to zero.

\vskip 3 true cm


\begin{thebibliography}{99}

\bibitem{Gotze} For reviews, see W. G\"otze, in {\it Liquids, freezing and 
glass transition}, Les Houches 1989, 
JP Hansen, D. Levesque, J. Zinn-Justin Editors,  North Holland. see also W. G\"otze, L. Sj\"ogren, {\it Rep. Prog. Phys.} {\bf 55} (1992) 241.

\bibitem{Science} For a review, see the interesting series of papers in {\it Science},
{\bf 267} (1995) 1924.

\bibitem{VF} H. Vogel, {\it Z. Phys. } {\bf 22} (1921) 645 ; G. S. Fulcher,
{\it J. Am. Ceram. Soc. }, {\bf 6} (1925) 339 

\bibitem{Bass} see the lucid papers of H. Bassler, {\it {\it  Phys. Rev. Lett.}} {\bf 58} (1987) 767
and R. Richter, H. Bassler, {\it J. Cond. Mat. Phys.} {\bf 2} (1990) 2273, 
and references therein.

\bibitem{Comments} for an enlightening introduction to the experimental controversy,
 see the series of Comments in {\it  Phys. Rev. E}: X.C. Zeng, D. Kivelson, G. Tarjus, Phys.
Rev. E {\bf 50} (1994) 1711, P. K. Dixon, N. Menon, S. R. Nagel, {\it  Phys. Rev. E} {\bf 50}
(1994) 1717, H. Z. Cummins, G. Li, {\it  Phys. Rev. E} {\bf 50} (1994) 1720, and references 
therein, in particular  H. Z. Cummins, W.M. Du, M. Fuchs, W. Gotze, S. Hildebrand, A.
Latz, G. Li, N.J. Tao,  {\it  Phys. Rev. E} {\bf 47} (1993)  4223

\bibitem{Thiru} T. R. Kirkpatrick, D. Thirumalai, P. G. Wolynes, 
{\it  Phys. Rev. A} {\bf 40} (1989) 1045
 and references therein.

\bibitem{CuKu} L. Cugliandolo, J. Kurchan, {\it  Phys. Rev. Lett.} {\bf 71} (1993) 173, 
L. Cugliandolo, P. Le Doussal, {\it Large time off-equilibrium dynamics of a particle
diffusing in a random potential}, preprint cond-mat 9505112, L. Cugliandolo, P. Le
Doussal, J. Kurchan, preprint cond-mat {\bf 9509008}. S. Franz, M. M\'ezard, Europhys.
Lett. {\bf 26} (1994) 209,  { Physica } {\bf A209} (1994) 1

\bibitem{us}  J.P. Bouchaud, L. Cugliandolo, J. Kurchan,  M. M\'ezard,
 {\it Mode Coupling Approximations, Glass Theory and Disordered Systems}, preprint
cond-mat {\bf 9511042}.

\bibitem{SIQD} J.P. Bouchaud, M. M\'ezard {\it  J. Physique I France} {\bf 4} (1994) 1109,
E. Marinari, G. Parisi, and F. Ritort,{\it  J. Phys. A} {\bf 27} (1994) 7615 ; 
{\it  J. Phys. A} {\bf 27} (1994) 7647, L. F. Cugliandolo,  J. Kurchan, G. Parisi, F.Ritort, {\it  Phys. Rev. Lett.} {\bf 74} (1995) 1012, 
R. Monasson, {\it  Phys. Rev. Lett.}
{\bf 75} (1995) 2847, P. Chandra, L. Ioffe and D. Sherrington; cond-mat {\bf 9502018}.
P. Chandra, M. Feigelmann and L. Ioffe;
preprint cond-mat  {\bf 9509022}. See also the interesting work by 
S. Obuhkov, D. Kobzev, D. Perchak, M. Rubinstein, and C. Reuner, H. L\"owen, J.L. Barrat on rotating hard rods (preprints, 1995).


\bibitem{Ritort} F. Ritort, {\it  Phys. Rev. Lett.} {\bf 75} (1995) 1190;
 S. Franz, F.Ritort, {\it
Dynamical solution of a model
without energy barriers}, preprint cond-mat {\bf 9505115}
to appear in {\it Europhys. Lett.}
C. Godr\`eche, J.P. Bouchaud, M. M\'ezard, to appear in {\it  J. Phys. A}, 
C. Godr\`eche, J.M. Luck, in preparation.

\bibitem{Vilgis} T. Vilgis,{\it  J. Phys. Cond. Matter}, {\bf 2} (1990) 3667,

\bibitem{Odagaki1} T. Odagaki, J. Matsui, Y. Hiwatari, 
{\it Physica A} {\bf 204} 464 (1994), and refs. therein.

\bibitem{Odagaki2} T. Odagaki, {\it  Phys. Rev. Lett.} {\bf 75} (1995) 3701.

\bibitem{BCM95} J.P. Bouchaud, A. Comtet and C. Monthus, cond-mat {\bf 9506027}, to appear in {\it  J. Physique I}(Dec. 1995).

\bibitem{Franz} S. Franz, J. Hertz, {\it  Phys. Rev. Lett.} {\bf 74} (1995) 2114

\bibitem{Bou92} J.P. Bouchaud, {\em {\it  J. Physique I} (France)}
{\bf 2},  (1992) 1705. see also:  A. Barrat and  M. M\'ezard; {\it J. Physique I (France) } {\bf 5} (1995) 941.

\bibitem{Struik} For references of aging in {\it mechanical} properties, see 
 L.C.E. Struick, ``Physical Aging in Amorphous
Polymers and Other Materials'' (Elsevier, Houston, 1978). It would be interesting to observe aging in more microscopic measurements, 
such as dynamical light scattering, etc.

\bibitem{rev} For a review on aging properties in experimental spin-glasses, see
 E. Vincent, J. Hammann, M. Ocio, p. 207 in
"Recent Progress in Random Magnets", D.H. Ryan Editor, (World
Scientific Pub. Co. Pte. Ltd, Singapore 1992).


\bibitem{BVH} J.P. Bouchaud, E. Vincent, J. Hammann, {\it  J.
Physique I}, 4, 139 (1994)

\bibitem{MPV} M. M\'ezard, G. Parisi, M.A. Virasoro, `Spin Glasses and Beyond', World
Scientific (1987).

\bibitem{Derrida} B. Derrida, {\it Phys. Rev. } {\bf B 24},
(1981) 2613 

\bibitem{RF} J.P. Bouchaud, A. Comtet, A. Georges, P. Le Doussal, {\it  Ann. Phys.} {\bf 201} (1990) 285, 
V.S. Dotsenko, M. V. Feigelmann, L.B. Ioffe, `Spin-Glasses and related problems', Soviet
Scientific Reviews, vol. 15 (Harwood, 1990).

\bibitem{BG90}
J.P. Bouchaud and A. Georges,  
{\em Phys. Rep.} {\bf 195} (1990) 127.

\bibitem{BM} J.P. Bouchaud, M. M\'ezard, in preparation.

\bibitem{BD95} J.P. Bouchaud, D.S. Dean,  {\it  J. Physique I (France)}
{\bf 5},  (1995) 265.

\bibitem{Bardou} F. Bardou, J. P. Bouchaud, O. Emile, C. Cohen-Tannouji, A.
Aspect, {\it  Phys. Rev. Lett.} {\bf 72} (1994) 203



\bibitem{Castaing} B. Castaing, J. Souletie, {\it  J. Physique I} (France) {\bf 1} (1991) 403.

\bibitem{DPS} W. Steffen, A. Patkovski, H. Gl\"aser, G. Meier, E.W. Fischer, 
{\it Phys. Rev.} {\bf E 49} (1994) 2992.

\bibitem{Arkipov} V. Arkhipov, H. Bassler, {\it  Phys. Rev. E} {\bf 52} (1995) 
1227 and references therein. Note that the authors suggests a rather low value for
$\nu \simeq 0.3$, which is not inconsistent with the behaviour of the stretched 
exponential exponent $\alpha$ reported in \cite{DPS}.

\bibitem{Souletie} J. Souletie, J. Appl. Phys. {\bf 75} (1994) 5512.

\bibitem{CTRW} for a review, see:  E. Montroll, M. Shlesinger, 
`On the Wonderful world of Random Walks', in Nonequilibrium
phenomena II, From stochastic to hydrodynamics, 
Studies in statistical mechanics 
XI (J.L. Lebowitz and E.W.
Montroll, eds) (North Holland, Amsterdam, 1984), 
M. Shlesinger, G. Zaslavsky, J. Klafter, {\it Nature }(London) {\bf 363} 31 (1993).

\bibitem{Japrl} S. Gomi and F. Yonezawa, {\it Phys. Rev. Lett.}  {\bf 74} (1995) 4125.



\end{thebibliography}
\end{document}